%
\let\useblackboard=\iftrue
%
%
\newfam\black
\input harvmac.tex

%
\input epsf.tex
\ifx\epsfbox\UnDeFiNeD\message{(NO epsf.tex, FIGURES WILL BE
IGNORED)}
\def\figin#1{\vskip2in}
\else\message{(FIGURES WILL BE INCLUDED)}\def\figin#1{#1}\fi
\def\ifig#1#2#3{\xdef#1{fig.~\the\figno}
\midinsert{\centerline{\figin{#3}}%
\smallskip\centerline{\vbox{\baselineskip12pt
\advance\hsize by -1truein\noindent{\bf Fig.~\the\figno:} #2}}
\bigskip}\endinsert\global\advance\figno by1}
\noblackbox
\def\Title#1#2{\rightline{#1}
\ifx\answ\bigans\nopagenumbers\pageno0\vskip1in%
\baselineskip 15pt plus 1pt minus 1pt
\else
\def\listrefs{\footatend\vskip
1in\immediate\closeout\rfile\writestoppt
\baselineskip=14pt\centerline{{\bf
References}}\bigskip{\frenchspacing%
\parindent=20pt\escapechar=` \input
refs.tmp\vfill\eject}\nonfrenchspacing}
\pageno1\vskip.8in\fi \centerline{\titlefont #2}\vskip .5in}
 
scaled\magstep3
 
scaled\magstep3
 
scaled\magstep3
 
scaled\magstep3
 
scaled\magstep3
\ifx\answ\bigans\def\tcbreak#1{}\else\def\tcbreak#1{\cr&{#1}}\fi

\useblackboard
\message{If you do not have msbm (blackboard bold) fonts,}
\message{change the option at the top of the tex file.}

\font\blackboard=msbm10 scaled \magstep1
\font\blackboards=msbm7
\font\blackboardss=msbm5
\textfont\black=\blackboard
\scriptfont\black=\blackboards
\scriptscriptfont\black=\blackboardss

\else

\fi
%
\def\yboxit#1#2{\vbox{\hrule height #1 \hbox{\vrule width #1
\vbox{#2}\vrule width #1 }\hrule height #1 }}
\def\fillbox#1{\hbox to #1{\vbox to #1{\vfil}\hfil}}
\def\ybox{{\lower 1.3pt \yboxit{0.4pt}{\fillbox{8pt}}\hskip-0.2pt}}

\def\comments#1{}

\def\half{{1\over 2}}

\def\CN{{\cal N}}

\def\II{\relax{I\kern-.07em I}}

\def\IIB{{\II}B}

\def\inbar{\,\vrule height1.5ex width.4pt depth0pt}
\def\IZ{\relax\ifmmode\mathchoice
{\hbox{\cmss Z\kern-.4em Z}}{\hbox{\cmss Z\kern-.4em Z}}
{\lower.9pt\hbox{\cmsss Z\kern-.4em Z}}
{\lower1.2pt\hbox{\cmsss Z\kern-.4em Z}}\else{\cmss Z\kern-.4em
Z}\fi}
\def\IB{\relax{\rm I\kern-.18em B}}
\def\IC{{\relax\hbox{$\inbar\kern-.3em{\rm C}$}}}
\def\ID{\relax{\rm I\kern-.18em D}}
\def\IE{\relax{\rm I\kern-.18em E}}
\def\IF{\relax{\rm I\kern-.18em F}}
\def\IG{\relax\hbox{$\inbar\kern-.3em{\rm G}$}}
\def\IGa{\relax\hbox{${\rm I}\kern-.18em\Gamma$}}
\def\IH{\relax{\rm I\kern-.18em H}}
\def\IK{\relax{\rm I\kern-.18em K}}
\def\IP{\relax{\rm I\kern-.18em P}}
\def\pp{{\relax{=\kern-.42em |\kern+.2em}}}

\font\cmss=cmss10 \font\cmsss=cmss10 at 7pt
\def\IR{\relax{\rm I\kern-.18em R}}

\def\frac#1#2{{{#1} \over {#2}}}

%
%

\def\NP{{\it Nucl. Phys.\ }}
\def\AP{{\it Ann. Phys.\ }}
\def\PL{{\it Phys. Lett.\ }}
\def\PR{{\it Phys. Rev.\ }}

\def\JHEP{{\it JHEP \ }}
\def\ATMP{{\it Adv. Theor. Math. Phys.\ }}
\def\CQG{{\it Class. Quantum Grav.\ }}
\lref\kallosh{R. Kallosh and A. Linde, ``Supersymmetry and the Brane
World,'' hep-th/0001071.}
\lref\chamblin{A. Chamblin and G.W. Gibbons, ``Supergravity on the
Brane,'' hep-th/9909130.}
\lref\girardello{L. Girardello, M. Petrini, M. Porrati and A. Zaffaroni,
``Novel Local CFT and Exact Results on Perturbations of N=4 Super Yang 
Mills from AdS Dynamics,'' \JHEP {\bf 9812} (1998) 022, hep-th/9810126.}
\lref\porrati{M, Porrati and A. Starinets, ``RG Fixed Points in
Supergravity Duals of 4-d Field Theory and Asymptotically AdS
Spaces,'' hep-th/9903085.}
\lref\girardellotwo{L. Girardello, M. Petrini, M. Porrati and
A. Zaffaroni, ``Confinement and Condensates Without Fine Tuning in
Supergravity Duals of Gauge Theories,'' \JHEP {\bf 9905} (1999) 026, hep-th/9903026.}
\lref\girardellothree{L. Girardello, M. Petrini, M. Porrati and
A. Zaffaroni, ``The Supergravity Dual of N=1 Super Yang-Mills
Theory,'' hep-th/9909047.}
\lref\hairreview{J.D. Bekenstein, ``Black Hole hair: twenty-five years 
after,'' hep-th/9605059; M. Heusler, ``No-Hair Theorems and Black
Holes with Hair,'' {\it Helv. Phys. Acta} {\bf 69} (1996) 501, gr-qc/9610019.}
\lref\sudarsky{D. Sudarsky, ``A simple proof of a no-hair theorem in
Einstein-Higgs theory,'' \CQG {\bf 12} (1995) 579; D. Sudarsky and
T. Zannias, ``Spherical Black Holes Cannot Support Scalar Hair,'' \PR
{\bf D58} (1998) 087502, gr-qc/9712083.}
\lref\cai{R.-G. Cai and J.-Y. Ji, ``Hairs on the cosmological
horizon,'' \PR {\bf D58} (1998) 024002, gr-qc/9708064.}
\lref\polchinski{J. Polchinski, ``Renormalization and Effective
Lagrangians,'' \NP {\bf B231} (1984) 269.}
\lref\gubsertwo{S.S. Gubser, ``Curvature Singularities: the Good, the Bad 
and the Naked,'' hep-th/0002160.}
\lref\sundborg{ P. Haggi-Mani and B. Sundborg, ``Free Large N
Supersymmetric Yang-Mills Theory as a String Theory,''
hep-th/0002189.}
\lref\gubserone{S.S. Gubser, ``Dilaton-driven confinement,''
hep-th/9902155.}
\lref\constable{N.R. Constable and R.C. Myers, ``Exotic Scalar States
in the AdS/CFT Correspondence,'' \JHEP {\bf 9911} (1999) 020, hep-th/9905081.}
\lref\kehagias{A. Kehagias and K. Sfetsos, ``On running couplings in
gauge theories from type IIB supergravity,'' \PL {\bf B454} (1999)
270, hep-th/9902125.}
\lref\odintsov{S. Nojiri and S.D. Odintsov, ``Two-Boundaries AdS/CFT
Correspondence in Dilatonic Gravity,'' \PL {\bf B449} (1999) 39,
hep-th/9812017.}
\lref\johnson{C.V. Johnson, A.W. Peet and J. Polchinski, ``Gauge
Theory and the Excision of Repulson Singularities,'' hep-th/9911161.}
\lref\myers{R.C. Myers, ``Dielectric-Branes,'' \JHEP {\bf 9912} (1999)
022, hep-th/9910053.}
\lref\klebanov{I.R. Klebanov and E. Witten, ``AdS/CFT Correspondence
and Symmetry Breaking,'' \NP {\bf B556} (1999) 89, hep-th/9905104.}
\lref\breitenlohner{P. Breitenlohner and D.Z. Freedman, ``Stability
in Gauged Extended Supergravity,'' \AP {\bf 144} (1982) 249.}
\lref\bekensteintwo{J.D. Bekenstein, ``Novel ``no-scalar-hair''
theorem for black holes,'' \PR {\bf D51} (1995) R6608.}
\lref\deboer{J. de Boer, E. and H. Verlinde, ``On the
Holographic Renormalization Group,'' hep-th/9912012.}
\lref\nastase{H. Nastase and D. Vaman, ``On the nonlinear KK
reductions on spheres of supergravity theories,'' hep-th/0002028;
 M. Cvetic, H. Lu, C.N. Pope and A. Sadrzadeh, ``Consistency of
Kaluza-Klein Sphere Reductions of Symmetric Potentials,'' hep-th/0002056.}
\lref\polyakov{S.S. Gubser, I.R. Klebanov and A.M. Polyakov, \PL {\bf
B428} (1998) 105, hep-th/9802109.}
\lref\freedman{D.Z. Freedman, S.S. Gubser, K. Pilch and N.P. Warner,
``Renormalization Group Flows from Holography - Supersymmetry and a
c-Theorem,'' hep-th/9904017.}
\lref\review{O. Aharony, S.S. Gubser, J. Maldacena, H. Ooguri and
Y. Oz, ``Large N Field Theories, String Theory and Gravity,''
hep-th/9905111v3.}
\lref\alvarez{E. Alvarez and C. Gomez, ``Geometric Holography, the
Renormalization Group and the c-Theorem,'' \NP {\bf B541} (1999) 441,
hep-th/9807226.}
\lref\wald{R.M. Wald, ``Gravitational Collapse and Cosmic
Censorship,'' gr-qc/9710068.}
\lref\christodolou{D. Christodoulou, {\it Ann. Math.} {\bf 140} (1994) 
607;  {\it Ann. Math.} {\bf 149} (1999) 183.}
\lref\distlertwo{J. Distler and F. Zamora, ``Chiral Symmetry Breaking in
the AdS/CFT Correspondence,'' hep-th/9911040.}
\lref\distlerone{J. Distler and F. Zamora, ``Non-Supersymmetric
Conformal Field Theories from Stable Anti-de Sitter Spaces,''
\ATMP {\bf 2} (1999) 1405, hep-th/9810206.}
\lref\igor{I.R. Klebanov and A.A. Tseytlin, ``Gravity Duals of
Supersymmetric $SU(N) \times SU(N+M)$ Gauge Theories,'' hep-th/0002159.}

\Title{ \vbox{\baselineskip12pt\hbox{hep-th/0003042}
\hbox{BROWN-HET-1218}
}}
{\vbox{
\centerline{Renormalization Group Flows }
\centerline{from Gravity in Anti-de Sitter 
Space}
\centerline{versus Black Hole No-Hair Theorems}}}

\centerline{ David A. Lowe}
\medskip

\centerline{Department of Physics}
\centerline{Brown University}
\centerline{Providence, RI 02912, USA}
\centerline{\tt lowe@het.brown.edu}
\bigskip

\centerline{\bf{Abstract}}

Black hole no-hair theorems are proven using inequalities that govern
the radial dependence of spherically symmetric configurations of
matter fields. In this paper, we analyze the analogous inequalities
for geometries dual to renormalization group flows via the AdS/CFT
correspondence. These inequalities give much useful information about
the qualitative properties of such flows. For Poincare invariant
flows, we show that generic flows of relevant or irrelevant
operators lead to singular geometries. For the case of irrelevant
operators, this leads to an apparent conflict with Polchinski's
decoupling theorem, and we offer two possible resolutions to this problem.

\vfill
\Date{\vbox{\hbox{\sl February, 2000}}}

\newsec{Introduction}

The conjectured correspondence between Type \IIB\ string theory on a $AdS_5
\times S^5$ background, and large $N$ $SU(N)$ maximally supersymmetric 
Yang-Mills theory, promises to offer new insights into conventional
quantum field theory (see \review\ for a general review). In the large 
$N$ limit, with strong 't Hooft coupling $g_{YM}^2 N$, 
the correspondence  provides us with a
description of the super Yang-Mills theory in terms of collective
coordinates, which are simply the fields of compactified Type \IIB\
supergravity. This leads to a new kind of strong coupling expansion, which has
the advantage that four-dimensional Poincare invariance is manifest.

Gauge theories with fewer or no
supersymmetry are of much greater physical interest. 
These may be studied by perturbing the couplings of the
maximally supersymmetric Yang-Mills theory at an ultraviolet scale,
and studying the renormalization group flow to the infrared. By the
AdS/CFT correspondence these RG flows are dual to asymptotically
anti-de Sitter geometries. The UV perturbation appears as boundary
conditions on the supergravity fields at large radius, and the radial 
dependence of the flow as one moves into the interior of the space  
encodes the RG flow to the IR.

It has been argued that flows involving just the relevant and marginal 
couplings of the Yang-Mills theory may be studied by truncating to
five dimensional gauged supergravity (see \review\ for a discussion
and further references). Numerous gravity duals of RG
flows have been constructed within this framework \refs{\girardello
\girardellothree \freedman \distlerone {--}\distlertwo}. 

The geometries involved have a close relation to the large mass limit
of black hole geometries. In fact, many of the examples in the previous
paragraph can be thought of precisely in this way. It is natural then
to ask what the analogs of the black hole no-hair theorems have to say 
about the geometries dual to Poincare invariant RG flows. This will be 
the main focus of this paper. 

Black hole no-hair theorems apply to asymptotically flat spherically
symmetric black holes
in theories of gravity coupled to scalar fields. Some helpful reviews
of this subject are \hairreview. Also see \refs{\sudarsky,
\bekensteintwo} for modern versions of the theorems. 
The theorems place conditions on the interaction potential of the
scalar fields: that it be positive semidefinite \sudarsky, or convex
\bekensteintwo. The theorems then proceed by deriving inequalities
which govern the radial dependence of the scalars, with the result
that scalar hair cannot depend on any continuous parameters at
infinity. These results have been generalized to spherically symmetric 
black holes in anti-de Sitter space in \cai. 

The scalar potential of compactified supergravity does not
satisfy the constraints used in the above theorems. The scalar
potential appearing in the five-dimensional gauged supergravity, for
example,  is not 
positive definite, and has directions where the potential asymptotes
to either $+\infty$ or $-\infty$. When we follow through the same
conditions in the context of the RG flow geometries, we will find hair 
is allowed, at least for finely tuned choices of the
perturbations. Nevertheless, the analogs of the inequalities on the
flow of the scalars will provide us with much useful qualitative
information about the properties of these flows, generalizing the
c-theorem of \refs{\alvarez,\freedman}.

For generic perturbations in the UV, we find the gravity dual of the
RG flow becomes singular in the interior. 
For relevant operators this is perhaps not
such a big surprise - from the field theory point of view, a generic 
perturbation will lead to a theory that is either free or trivial
(confining) in the infrared. Since gravity is inherently a non-linear
theory, the only way it can accommodate this is for the geometry to
become singular. For irrelevant operators, on the other hand, this is
more troubling, as we find that the singularity will depend in detail
on the UV perturbation of the irrelevant couplings, of which we have
an infinite number as $N \to \infty$. At first sight,
this looks like a violation of Polchinski's theorem \polchinski, which
states that all couplings in the infrared should be determined by the
finite number of
relevant and marginal couplings alone.

We offer two possible explanations of this fact:
either a large class of irrelevant operators in the super Yang-Mills are
dangerous in the sense that they become relevant along a typical RG
flow; or the dependence of the supergravity solution on the scalar
dual to an irrelevant operator is a red
herring as far as the infrared physics goes, and this dependence goes away
when the stringy resolution of the singularity is properly understood.
The first possibility seems to indicate the large $N$ Yang-Mills theory is sick
in the sense we have a massive failure of the decoupling of irrelevant 
operators from low energy physics. The second possibility is more
desirable from the quantum field theory point of view, 
but indicates some spectacular
stringy miracles are needed to properly understand the infrared physics.

\newsec{Inequalities for RG flows in AdS/CFT}

For gravity duals of Poincare invariant RG flows, the
five-dimensional asymptotically anti-de Sitter part of the
spacetime will have a metric of the form 
\eqn\metric{
ds^2 = e^{2 A(r)} ( dx^\mu)^2 - dr^2~,
}
where $\mu=0,1,2,3$.
We will be interested in solutions where this metric couples to scalar 
fields, with an action of the form
\eqn\scalarac{
S_\phi = {1 \over 2 \kappa_5^2} \int d^5x \sqrt{|g|} \biggl( -R + \half g^{\alpha 
\beta} \partial_\alpha \phi \partial_\beta \phi -  V(\phi) \biggr)~.
}
The generalization to multiple scalar fields is immediate, assuming
the kinetic term is positive definite.
We will adopt conventions where $\kappa_5=1$, and choose the negative
cosmological constant (the constant term in $V$) so that $A =r$ is
the vacuum solution as
$r\to \infty$.

We wish to view the renormalization group flow in the Wilsonian sense, 
so rather than trying to introduce counter-terms and sending the UV
cutoff to infinity, we will work with an explicit UV cutoff. Thus we will
impose boundary conditions on the scalar fields at some fixed, large
value of $r=r_{UV}$. Near this value of $r$ we will require that the
scalar field perturbation is small, so that the vacuum Einstein
equations hold (i.e. $A \sim r$).

There are two independent boundary conditions possible for scalar
fields: 
\eqn\boundc{
\phi =
\phi_+ \exp(\alpha_+ r)+ \phi_- \exp(\alpha_- r)~.
}
where $\alpha_\pm = -2 \pm 2 \sqrt{1 +{m^2\over 4}}$, for a scalar of mass
$m$. For
relevant operators $m^2 <0$,  but satisfies a unitarity bound
\breitenlohner\ provided $m^2 > -4$. This bound guarantees that the energy
of fluctuations of the field is positive definite.  
Marginal operators have $m^2=0$, while irrelevant operators
have positive $m^2$. Examples of irrelevant operators with finite
conformal dimensions are chiral primaries dual to massive
Kaluza-Klein modes of Type IIB supergravity on $AdS_5 \times
S^5$. Stringy modes will typically have conformal dimensions that
diverge in the large $N$ limit, \polyakov.

In order to deform the Lagrangian of the Yang-Mills theory in the UV by a
coupling dual to one of 
these modes, the $\phi_+$ part of the solution should be the dominant
contribution
at $r=r_{UV}$. This is to ensure the correct two-point correlation function is
reproduced, as explained in detail in \klebanov. There they also point 
out for conformal dimensions $\Delta$ between 2 and 3, either possibility is
allowed, corresponding to inequivalent quantizations. This subtlety
will not be a concern for us, as we will be mostly interested in
deformations by irrelevant operators $\Delta >4$. 
Turning on the $\phi_-$ boundary condition corresponds
to considering a non-trivial vacuum state in the strong coupling large $N$
Yang Mills theory, which gives rise to expectation values for
operators dual to the gravity mode.\foot{The mapping of $\phi_+$ and
$\phi_-$ to perturbations and expectation values, respectively, in the 
field theory dual is precise only when the cutoff $r_{UV} \to
\infty$. For finite $r_{UV}$ it is not known in general how to make a precise
distinction between the two. We comment further on this point in the
following section.} 
The gravity solutions found in
\refs{\odintsov, \kehagias, \gubserone, \constable} correspond to flows of this type.

In quantum field theory, perturbations by irrelevant couplings 
drastically alter the UV behavior 
of the theory. For us,
this shows up as the large $r$ asymptotics of the spacetime
being altered, if we extrapolate past $r_{UV}$. 
However by imposing our boundary conditions at finite
$r$ there will be a scaling regime where for small perturbations the
geometry will be approximately AdS near the boundary $r_{UV}$.
It will not be possible
to remove the UV cutoff, keeping the coefficient of the irrelevant
coupling fixed. Of course this is precisely the situation in the usual 
Wilsonian renormalization group. 

Following Sudarsky's proof of the black hole no-hair theorems
\sudarsky\ we will now derive an equality the scalar solution must
satisfy for any RG flow.
Take $\phi = \phi(r)$, and consider the equation of motion of the scalar
\eqn\scaleom{
e^{-4 A} ( e^{4 A} \phi' )' =  {\partial V \over \partial \phi}~,
}
where $'$ denotes a derivative with respect to $r$.
Multiply this by $\phi'$, and define the ``energy'' 
\eqn\energy{
E= \half (\phi')^2 -  V(\phi)-6~,
}
to obtain
\eqn\sseom{
E' = - 4 A' (\phi')^2~.
}
The $-6$ in \energy\ simply shifts our definition by an irrelevant
constant for convenience.

Now the c-theorem of \freedman\ implies $A'' \leq 0$ for all $r$, provided
$\phi$ satisfies a version of the weak energy condition. Thus if
$A'>0$ at $r=r_{UV} $, $A'>0 $ for all $r<r_{UV}$. Therefore our
general inequality for scalar perturbations is
\eqn\inequal{
E' \leq 0~,
}
for all $r<r_{UV}$. 

Next we examine the boundary conditions on the fields to check whether 
non-trivial flows can exist which are compatible with \inequal. For
the tachyonic scalars \inequal\ is easy to satisfy. In that case we
can send $r_{UV}$ to infinity in a smooth way, so we know 
$E_{UV}=0$. The condition \inequal\ then implies that in the infrared
$E_{IR} >0$. Even if $\phi' \to 0$ in the infrared, this condition can
still be satisfied because $V$ is not bounded below. Thus these
solutions can display nontrivial hair. The solutions found in 
\refs{\girardello
\girardellothree \freedman \distlerone {--}\distlertwo} are of this
type. 

For massive scalars the potential is bounded below (ignoring for the
moment higher non-linear couplings to lighter fields, which should be
valid for sufficiently small perturbations). However now we are not
able to take our boundary $r_{UV}$ off to infinity. Instead we find
$E_{UV} = -2 \alpha_+ \phi_+^2 \exp(2 \alpha_+ r_{UV})<0$.  The
condition \inequal, implies $E_{IR}>E_{UV}$ which again is possible to 
satisfy for some solutions, so once again nontrivial hair is
possible. 

On the other hand, if we impose that the massive scalar field approach the
$\phi_-$ solution in the $UV$, we are free to scale $r_{UV}$ to
infinity. Now $E_{UV}=0$, and at least if $\phi$ decouples from the
other scalars in the potential, 
and starts out at a global minimum at infinity, 
the only possible solution is to have
$\phi$ be constant. Thus no hair is possible, which implies that
states in the Yang-Mills theory with expectation values turned on for
such irrelevant operators will not have good gravity duals. Of course such
Poincare invariant vacuum states are certainly not present in finite
$N$ Yang-Mills theory, so this is a nice consistency check.

To study the nontrivial solutions further, we will need some linear
combinations of the Einstein equations
\eqn\einstein{
\eqalign{
G_r^r &= \kappa_5^2 T_r^r  \qquad \Rightarrow \qquad
- 6 (A')^2 = - \half (\phi')^2 + V(\phi)~,  \cr
G_t^t - G_r^r &= \kappa_5^2 (T_t^t-T_r^r) \qquad \Rightarrow \qquad
-3 A'' =  (\phi')^2~. \cr}
}
The c-theorem of \freedman\ follows immediately from the last of these 
equations. Solving the first of these for $A'$ and taking the positive 
root, as required by the c-theorem, and inserting into \scaleom\
yields
\eqn\harmon{
\phi'' + {4\over \sqrt{6}} \phi' \sqrt{ \half (\phi')^2 -V(\phi)} =
{\partial V \over \partial \phi}~.
}
The term in the square root is just $E(r)+6$, so by \inequal\ this is
always positive along the flow, provided our UV perturbations are
small.
The equation \harmon\ now has a nice physical interpretation. It is
simply the equation governing a particle moving in the potential $-V$
with ``negative'' position dependent damping as we flow from the UV to
the IR. We can therefore immediately conclude that for generic
boundary conditions the negative damping will cause the solution to
become singular in the infrared. Fine tuning of the initial
perturbation in the UV will be needed to land
the solution on an extremum of the potential, to obtain a supergravity 
solution that smoothly interpolates from the UV to the IR. 

For the case of relevant operators examples of such singular flows
have been constructed in \refs{\girardello
\girardellothree \freedman \distlerone {--}\distlertwo}. The
singularities of such flows exhibit a certain universality as
emphasized in
\girardellotwo. Namely, for a wide range of parameter choices, the
potential becomes irrelevant near the singularity, which appears at
some finite radius $r=r_0$. Solving \harmon\
and \einstein\ yields
\eqn\singu{
\phi \sim \pm \sqrt{3 \over 4} \log(r-r_0) ~, \qquad 
A \sim {1\over 4} \log(r-r_0)~,
}
near the singularity. This singularity leads to a curvature
singularity both in the string and the Einstein frame metrics, from
the five and ten-dimensional points  of view.
For multiple scalar fields, one obtains the same 
behavior, with $\phi^a \sim n^a \log(r-r_0) $ for
some suitably normalized vector $n^a$, whose direction will depend in
detail on the initial perturbation in the UV. 

The same statements will 
apply also for the case of flows from perturbations by a large class
of irrelevant operators, because only the scalar kinetic term is
relevant in this limit. In general, it is easy to
find regions in the parameter space of initial conditions where 
 the $n^a$ will
depend in detail on the irrelevant
couplings in the UV.

In certain regimes of the parameter space of the UV perturbations, a
different kind of singularity may arise where the potential of the
scalar fields dominate the flow. This is always the case for flows
which preserve supersymmetry.
Such flows will depend on the details 
of the potential in the highly non-linear regime, so we will not
comment on them further here.

\ifig\fone{The scalar $\phi(r)$ dual to an irrelevant
coupling, together with a scalar $\rho(r)$ dual to a relevant
coupling. }
{\epsfysize=2in\epsfbox{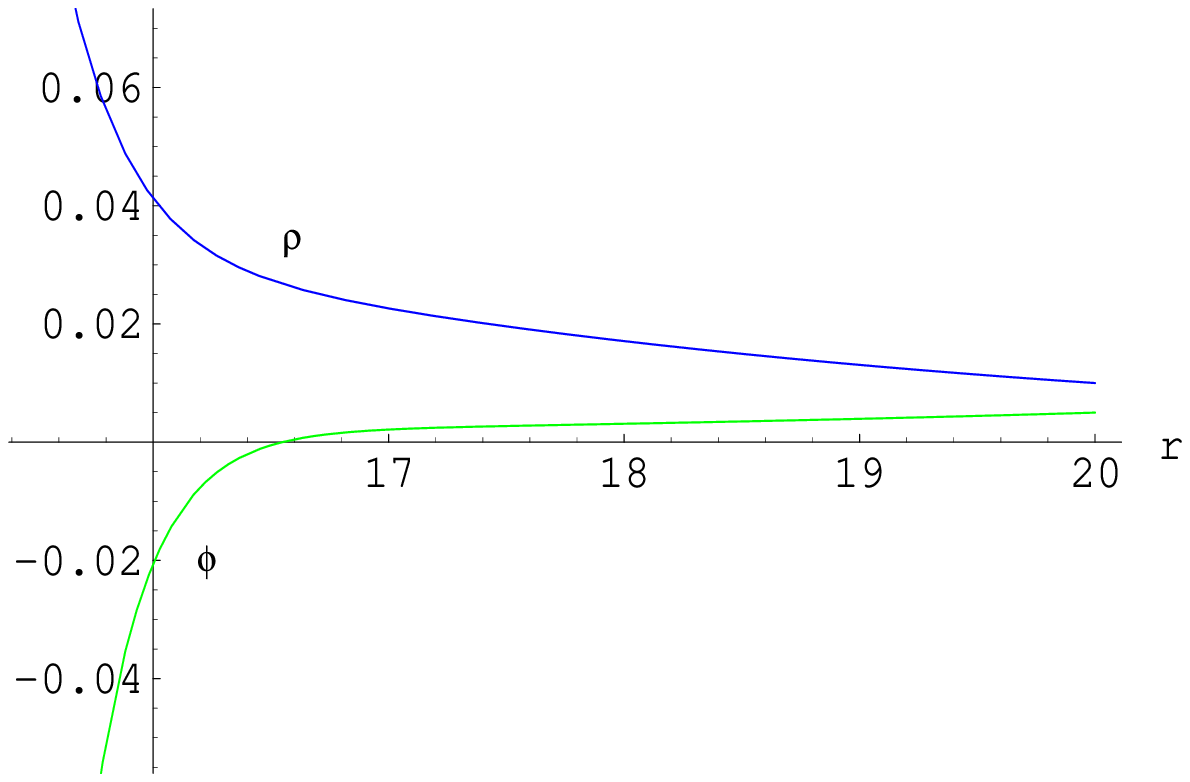}}

\ifig\ftwo{On the left, the
trajectory in the $E$, $\phi$ plane, together with the potential
$-V$. On the right, the corresponding trajectory in the $E$, $\rho$
plane.}{\epsfysize=2in\epsfbox{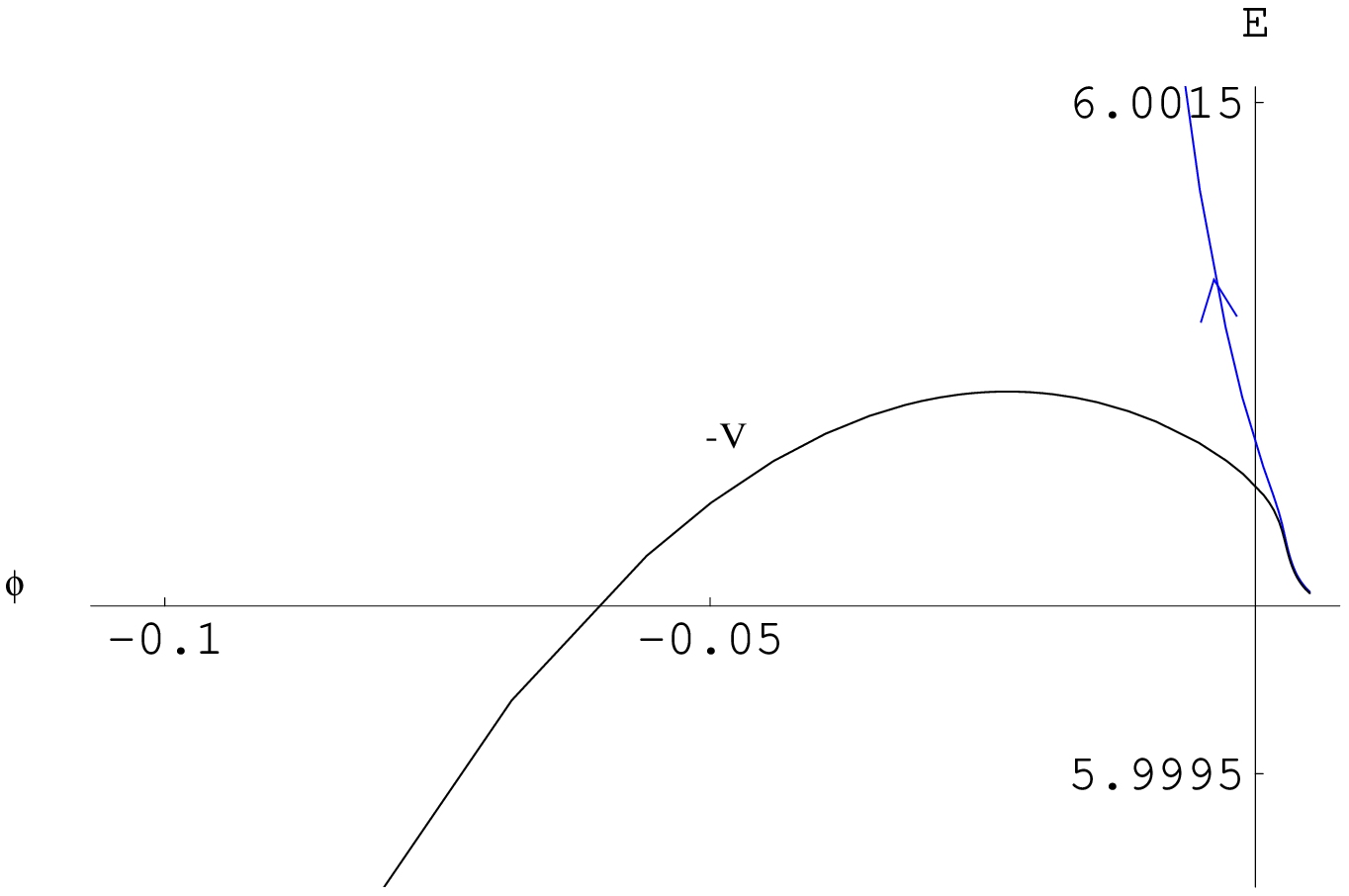}\epsfysize=2in\epsfbox{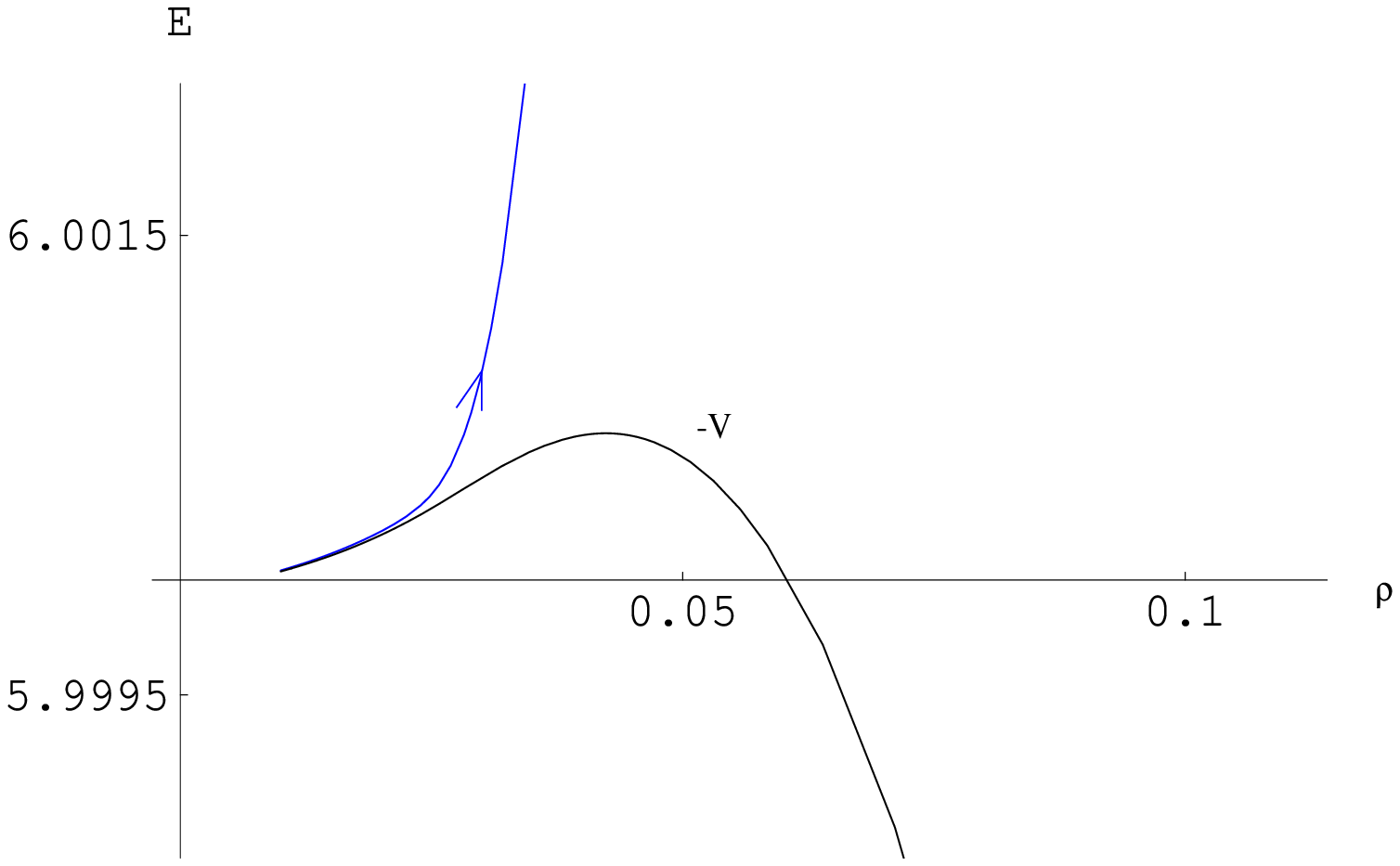}}

Let us now consider a concrete example of a flow of an irrelevant
operator dual to a massive Kaluza-Klein mode $\phi$ coupled to a relevant
operator, dual to a tachyonic mode $\rho$.
Unfortunately
the non-linear couplings of the massive Kaluza-Klein modes in the
$AdS_5\times S^5$ compactification of Type \IIB\ supergravity are not
known. See \nastase\ for recent developments in this direction. 
As a toy model for the flow, 
take the potential to be $V= \half \phi^2-\half \rho^2 -6$, and ignore the
non-linear couplings not already present in \harmon. Boundary
conditions are chosen at $r=r_{UV}$ so that $\phi_-=\rho_-=0$ and
$\phi_+,~ \rho_+ \ll 1$. The results of the numerical evolution toward the
infrared are shown in \fone\ and \ftwo. The singular behavior is precisely of the 
form \singu, (on the $+$ branch).

By making the perturbations sufficiently small, we can obtain an
arbitrarily large scaling region, where the relevant coupling is
increasing toward the infrared, while the irrelevant coupling is
decreasing. However when the relevant coupling dual to $\rho$ becomes
sufficiently large, it pushes both $\rho$ and $\phi$ into the runaway
behavior \singu. For reasonable choices of parameters, $\phi$ is of
the same order as $\rho$ near the singularity. 

If $m^2$ of the mode $\phi$ is taken to be large (e.g. we considered
the same example for $m^2=100$), the irrelevant mode
decreases more rapidly along the flow. However, because the $\rho$
field is monotonically increasing as $r$ decreases, eventually the
coupling to $\phi$ induces a mixing of the IR stable solution  of the
linearized equation with the IR unstable solution. For large $m^2$ the 
unstable mode increases correspondingly rapidly. Numerical evidence
points to the fact that the $n^a$ characterizing the runaway behavior
\singu, are typically dominated by the participating fields with the largest
value of $m^2$.

\newsec{Discussion}

For perturbations by relevant couplings we expect a large change 
in the supergravity
solution as we flow to the IR. 
However if the solution becomes singular for generic
perturbations it is still a bit puzzling. 
Of course from the field theory point of view it is natural that we
should need to fine tune couplings in the UV to land on a non-trivial
conformal field theory in the IR. A generic relevant perturbation, on
the other hand, will
cause the theory to flow to either a free fixed point in the IR, or a
trivial theory if all the physical degrees of freedom become massive. 
At least for free fixed points,
gravity will have a hard time reproducing the behavior as it is 
inherently non-linear. The only way for the supergravity solution to
circumvent this fact is for the solution to become singular. The
singularity associated with free fixed points
may be resolved by string theory in the tensionless limit as suggested 
in \sundborg, for the case of free $\CN=4$ Yang-Mills. 

For perturbations by irrelevant couplings the generically singular behavior
is rather more surprising from the point of view of
renormalization group
flow. One expects a small perturbation by an irrelevant operator in
the UV to produce a small change in the IR (assuming that operator
does not flow to become relevant). 
The precise statement is summarized by Polchinski's theorem \polchinski, which
says that along a renormalization group flow down to some scale
$\Lambda$, the unrenormalizable couplings at the scale $\Lambda$ are
determined in terms of the renormalizable couplings at $\Lambda$. The
flow is to a stable surface specified by the finite number of
renormalizable couplings. On the other hand, from the
gravity point of view, the flow to the IR
depends on the value of a potentially infinite
number of non-renormalizable couplings for $N\to \infty$, 
apparently contradicting
Polchinski's theorem. 

One possible resolution of this problem is as follows. The large $N$
Yang-Mills theory at strong coupling is plagued with an abundance of
dangerous irrelevant operators, which become relevant along a generic
renormalization group flow.\foot{An example of a full
ten-dimensional supergravity dual of such a flow can be found in
\igor.} While this behavior is consistent with
Polchinski's theorem, it implies we will have a hard time decoupling
high dimension operators from low energy physics. This rather limits
the utility of truncations to five dimensional gauged supergravity,
which describes only the relevant and marginal couplings in the UV.

Another possible resolution is the following. To
understand the correct IR physics we need to be in a region where
curvatures become of order the string scale. 
String theory resolves these curvature singularities by
correctly matching onto the highly non-universal supergravity
solution, but all the universal features of the flow to the IR 
are captured only
in the region where stringy effects are large. 
This seems like a tall
order, as it requires that all the low energy observables be
independent of the details of the supergravity solution in the
region of sub-stringy curvatures in the vicinity of the singularity. 
However there are examples where behavior of this type is at least
partially realized. If this possibility is realized, it seems to imply 
we learn little about the IR physics of large $N$ super Yang-Mills by
studying supergravity solutions alone.

One such example is extracting glueball masses from these
singular geometries. In \constable\ this calculation is performed for
glueballs dual to axion modes. The axion, and a large class of other
modes, see a repulsive potential as the singularity is approached. The 
mass spectrum of glueballs dual to these fluctuations is independent
of the details of the singularity. However, in the example considered
in \constable\ it is not hard to find other modes which see no
repulsive potential. The dilaton is one example. Without a resolution
of the singularity it is not possible to compute the spectrum of such
fluctuations. 

It is difficult to make general statements about how string theory may 
resolve these singularities. One class of singularities that have been 
studied in detail are the repulson
singularities that arise in D-brane configurations dual to 
large $N$ $SU(N)$ $\CN=2$ supersymmetric Yang-Mills theory \johnson. Here a
naked singularity is resolved by the formation of a smooth
distribution of D-brane charge. The basic idea here is likely to
generalize to a large class of singularities - brane charge spreads out,
giving a smooth configuration (see also \myers\ for studies of
this effect). From the supergravity point of view explicit sources are 
required in the supergravity equations of motion to represent the
presence of the D-branes. If stringy resolutions of this type are
relevant for the singularities we have discussed here, 
it supports the viewpoint that stringy
effects mask the IR physics from the dependence on the irrelevant
coupling parameters. 

One should also bear in mind the additional parameters in the
supergravity solutions corresponding to expectation values of
operators. As mentioned above, the mapping between boundary conditions 
at $r=r_{UV}$ and couplings and expectation values in the field theory 
becomes difficult for the singular geometries under consideration
here. As far as we know, the finite $N$ CFT does not possess a moduli space
of vacua corresponding to expectation values of non-marginal
operators. 
The supergravity dual of such a configuration, 
on the other hand, appears to be
well-defined away from the vicinity of the singularity. This suggests
that once the string theory resolution of these singularities is
understood, a large class of these singularities will be found to be
unstable. Presumably the stability of the singularity will provide the 
extra information needed to uniquely determine the boundary conditions 
at $r=r_{UV}$ corresponding to pure coupling perturbations.
We emphasize this problem cannot be analyzed within the
framework of low-energy supergravity alone, as in general boundary
conditions must be imposed at the singularity to determine the 
future evolution. A less attractive possibility is that these extra
``moduli'' are an artifact of the large $N$ limit.

One may wonder what the cosmic censorship conjecture has to say about
these singular solutions. See \wald\ for a review of this topic. The weak
cosmic censorship conjecture roughly states that for generic smooth
initial data, 
in asymptotically flat space, with suitable matter, any singularities
that form are necessarily hidden behind event horizons. There is much
nontrivial evidence for this conjecture, and no clear
counter-examples. 
However, there is
strong evidence that for spherically symmetric collapse of scalar
fields in asymptotically flat spacetime this weak cosmic censorship 
conjecture holds \christodolou. 
One difference in 
our case is the negative cosmological constant. The
stress-energy tensor (including the cosmological constant term)
violates the dominant
energy condition by allowing negative energy densities. This inhibits 
the formation of black hole horizons, permitting naked singularities
to appear in the evolution of scalar fields.  For us, perhaps the most 
important difference is that
four-dimensional Poincare
invariance of the solutions violates 
the condition that the initial data be generic. 

One might worry
that the boundary conditions for irrelevant couplings correspond to
infinite energies at infinity. However we insist on placing these
boundary conditions at finite radius. These solutions might then be
viewed as the interiors of smooth solutions which asymptote to anti-de 
Sitter space. Specifying the exterior solution corresponds in 
field theory language to specifying counter-terms in the bare action 
needed to keep
physical quantities finite as the cutoff is removed. 

It is interesting to consider the finite temperature version of the
inequality \inequal, and to test whether the singular solutions we
have described are stable with respect to turning on this additional
parameter. 
We can repeat the argument of the previous section for the 
metric
\eqn\bhmetric{
ds^2 = e^{2 A(r)} ( \mu(r) dt^2 - (dx^i)^2 ) - {dr^2 \over \mu(r)}~,
}
where $i=1,2,3$ and $\mu \sim 1 - M \exp(-4 r)$ at large $r$.
The scalar field equation motion now becomes
\eqn\bhseom{
(\mu \phi')' + 4 A' \mu \phi' =  {\partial V \over \partial \phi}~.
}
Multiplying by $\phi'$ and rearranging gives
\eqn\bhstuff{
\half (\mu (\phi')^2 )' + (\half \mu' + 4 A' \mu) (\phi')^2 = 
{\partial V \over \partial \phi} \phi'~.
}
We can now define an energy function
\eqn\bhen{
E = \half \mu (\phi')^2 -  V-6~,
}
which then satisfies, using \bhstuff\
\eqn\bhenbound{
E' = - ( \half \mu' + 4 A' \mu) (\phi')^2~.
}
Using the Einstein equations, it follows that each term on 
the right hand side
of \bhenbound\ is negative outside the horizon. Thus we obtain the
same condition for the new $E$ \inequal.
 
At a regular black hole horizon,
$\phi'$ will be finite, while $\mu$ will vanish. Thus $E= -
V(\phi_{hor})-6$ at the horizon. The conditions on $E$ near
the boundary are the same as before, namely for a relevant
perturbation we can assume $E_{UV}=0$, or for a irrelevant
perturbation $E_{UV}$ is small and negative. A smooth black hole will
exist provided $V(\phi_{hor})+6 < -E_{UV}$, which is easy to satisfy
for the tachyonic relevant modes, and possible to satisfy for
irrelevant modes. 

However we do not see any sign that the singular solutions we have
considered at zero temperature ($M=0$) will, in general,
 become hidden behind event 
horizons when we turn on $M$. To answer this question in more detail, 
we have numerically solved the Einstein
equations coupled to a scalar generalizing the example of
\fone\ to finite $M$, and found that the singularity persists until $M$ becomes
larger than some finite value. The singularities are therefore
stable against turning on an additional parameter, which points toward 
the fact that they are indeed physically meaningful.

While this manuscript was being prepared \gubsertwo\ appeared, where
very similar methods are used with a rather different emphasis.

\bigskip
{\bf Acknowledgments}

I wish to thank A. Jevicki, R. Myers, S. Ramgoolam and especially
Radu Tatar for helpful discussions.
This research is supported in part by DOE grant DE-FE0291ER40688-Task A.     
 
\vfil
\eject

\listrefs
\end